\documentclass[12pt,a4paper]{article}  

\usepackage{FDiagrams}

\def\tr{{\textrm{tr}}}
\def\Tr{{\textrm{Tr}}}
\def\T{{\textrm{T}}\,}

\newcommand{\be}{\begin{equation}}
\newcommand{\ee}{\end{equation}}

\newcommand{\ba}{\begin{equation} \begin{aligned}}
\newcommand{\ea}{\end{aligned} \end{equation}}

\def\la{\langle}
\def\ra{\rangle}

\newcommand{\pa}{\partial}

\def\del{\Delta}
\def\ddel{{}^\bullet\! \Delta}
\def\deld{\Delta^{\hskip -.5mm \bullet}}
\def\dddel{{}^{\bullet \bullet} \! \Delta}
\def\ddeld{{}^{\bullet}\! \Delta^{\hskip -.5mm \bullet}}

\begin{document}

\begin{center}
{\LARGE \bf  
Path integral calculation of heat kernel traces\\[3mm]  
with first order operator insertions}
\vskip 1.2cm
Fiorenzo Bastianelli$^{\,a,b}$  and Francesco Comberiati$^{\,a}$ 
\vskip 1cm

$^a${\em Dipartimento di Fisica e Astronomia, Universit{\`a} di Bologna,}

$^b${\em   INFN, Sezione di Bologna,}

 {\em   via Irnerio 46, I-40126 Bologna, Italy}
 
  \end{center}
\vskip .8cm

\abstract{We study generalized heat kernel  coefficients, which appear in the trace of the heat kernel 
with an insertion of a first-order differential operator, by using a path integral representation. 
These coefficients may be used to study gravitational anomalies, i.e. anomalies in the conservation of the stress tensor. 
We use the path integral method to compute  the coefficients related to the gravitational anomalies 
of theories in a non-abelian gauge background and flat space of dimensions 2, 4, and 6. 
In 4 dimensions one does not expect to have genuine gravitational anomalies.
However, they may be induced at intermediate stages by regularization schemes that fail to preserve 
the corresponding symmetry. A case of interest has recently appeared in the study of the
trace anomalies of Weyl fermions.
}

\section{Introduction}
Heat kernel methods provide a useful tool for investigating QFTs. They were introduced by Schwinger for studying 
QED processes \cite{Schwinger:1951nm} and extended to curved spaces and non-abelian gauge fields by 
DeWitt \cite{DeWitt:1965jb}. There are many reviews and books dedicated to them, as 
\cite{Barvinsky:1985an, DeWitt:2003pm, Vassilevich:2003xt}. 

One application of the heat kernel finds its place in the study of anomalies. 
The connection is most easily seen by recalling Fujikawa's method \cite{Fujikawa:1979ay}, which identifies 
the anomalies as arising from the non-invariance of the path integral measure under the symmetry transformations. 
In that approach the anomalies are cast as  regulated infinitesimal jacobians, 
$\lim_{\beta\to 0} {\rm Tr}\, J\, e^{-\beta {\cal R}}$, where $J$ is the generator of the anomalous symmetry 
and ${\cal R}$ a regulator, usually a second-order differential operator.  
Once Wick-rotated to euclidean space the regulator ${\cal R}$ becomes an elliptic operator and $e^{-\beta {\cal R}}$ 
defines the associated   heat kernel.
Often the operator $J$  depends only on the spacetime coordinates, and it does not contain any differential operator.
This is the case of the usual chiral and trace anomalies. 

Our interest in this paper is in traces that contain a first-order differential operator. This situation arises when 
one considers gravitational anomalies \cite{AlvarezGaume:1983ig}. The latter are  anomalies in the conservation 
of the stress tensor, and the corresponding symmetry is the arbitrary change of coordinates (diffeomorphisms).
Diffeomorphisms are generated by the Lie derivative of the quantum fields, and  on scalars and  Dirac spinors
the Lie derivative ${\cal L}_\xi $ takes the simple form
\be
{\cal L}_\xi =\xi^\mu(x) \partial_\mu
\ee 
where $\xi^\mu(x)$ is the vector field due to  an infinitesimal change of coordinates ($x^\mu \to x'^\mu = x^\mu - \xi^\mu(x)$).
The corresponding anomaly is then related to the regularization of an infinitesimal Fujikawa jacobian of the form
\be
J = \Big [\xi^\mu(x) \partial_\mu + \sigma (x)\Big]\delta^D(x-y)
\label{J}
\ee
where $\sigma(x)$ is a function that depends on the regularization scheme adopted (often one   takes 
$\sigma (x) = \frac12 \partial_\mu \xi^\mu(x) $ as a convenient choice, see \cite{Fujikawa:2004cx, Bastianelli:2006rx}).
An application of this type of traces has recently appeared in \cite{Kurkov:2018pjw}.
 
 In this paper,  we shall study  traces of the heat kernel with an insertion of a first-order differential operator
 of the form given in   \eqref{J}  by using a quantum mechanical path integral.
 After making explicit  the relationship between the  heat kernel traces and their path integral representation, 
 we use the latter to evaluate the first three heat kernel  coefficients for an elliptic operator ${\cal R}$ 
 containing  a non-abelian gauge field $A_\mu(x)$ and an arbitrary matrix-valued scalar potential $V(x)$.  
These coefficients are a generalization of the standard heat kernel coefficients, also  known  as Seeley-DeWitt 
coefficients, as they contain the insertion of a first-order differential operator. 
We shall call them generalized heat kernel coefficients, for simplicity.
 Some of these coefficients have  been computed before in \cite{Bastianelli:1990xn} and  \cite{Branson:1997ze}.
Here we shall reproduce some of those results with the path integral method, and compute an additional one. 

Our motivation for investigating these coefficients stems from a desire of addressing the anomalies of a
Weyl fermion in four dimensions by using a regularization scheme that induces gravitational anomalies as well. 
 This situation appeared in \cite{Bastianelli:2018osv}, where the trace anomaly of a Weyl fermion in 
 an abelian gauge background was computed to verify the absence of a parity violating term, 
 conjectured in  \cite{Nakayama:2012gu} to be a possibility for CP violating theories. 
The use of a Pauli-Villars (PV) regularization with a Majorana mass showed the absence of such a term, as  the PV Majorana mass 
 preserves CP and  diffeomorphism invariance. On the other hand, the verification of  the same result with
a PV Dirac mass could not be completed,  as the latter induces gravitational anomalies, which can  
be computed by using a generalized heat kernel coefficient
in the background of a non-abelian gauge field and flat spacetime. 
This justifies the use of flat space that we consider 
in our analysis. The non-abelian background is however
needed as the regulator $\cal R$ 
contains gamma matrices, making the connection contained in $\cal R$ effectively  non-abelian, 
even for the simple case of a Weyl fermion coupled to a $U(1)$ gauge field.

Thus, the problem we face in this paper is to study the path integral method
to compute generalized  heat kernel coefficients, verifying the ones previously known and producing a new one.
In section \ref{one} we review the path integral representation of the heat kernel and its traces.
 We start by considering a simple elliptic operator $\cal R$, interpreted 
as the quantum hamiltonian of a non-relativistic particle in a scalar potential, and study 
how to insert an arbitrary function of the particle coordinates inside the heat kernel trace.
We discuss the role played by the propagators defined either by the Dirichlet boundary conditions (DBC)  or by 
the string inspired (SI) method, which can be used equivalently for generating the perturbative expansion of the 
path integral. Section \ref{two} extends the previous set-up to include the insertion of a first-order differential operator
inside the heat kernel trace, and uses a more general hamiltonian  $\cal R$ 
with a non-abelian gauge potential $A_\mu$ and a matrix-valued scalar potential $V$.
The corresponding particle action is also matrix-valued, and  the path integral contains 
a time ordering prescription to maintain gauge covariance.
In section \ref{three} we present the first three generalized heat kernel coefficients, 
and in  section \ref{four} we describe the calculation of the simplest one
with the path integral method, reproducing  the result of \cite{Bastianelli:1990xn}.
Having verified the consistency of the method, in section \ref{five}  we proceed 
with the calculation of two more heat kernel coefficients. 
The first one is the flat space limit of a more general result originally obtained
in \cite{Branson:1997ze}, which is relevant for the gravitational anomalies in a flat four-dimensional space.
The last coefficient is new, and may be useful for the gravitational anomalies of gauge theories 
in a flat six-dimensional space. After our conclusions, 
 we report in appendix \ref{A} the worldline propagators defined by the Dirichlet boundary conditions
   and by the string inspired method, in appendix \ref{B} we use them  for
 computing some simple Seeley-DeWitt coefficients  as a simple review of the path integral method, and  
in appendix \ref{C} we report further calculational details.

\section{Path integral representation of heat kernel traces}
\label{one}

The path integral  representation was used to study the trace anomalies 
in \cite{Bastianelli:1991be} and \cite{Bastianelli:1992ct}, where the object of interest 
was represented by a heat kernel trace of the form
\be \label{eq:3}
{\rm Tr} \left[ \sigma(x)\, e^{-\beta {\cal R}}\right] 
\ee
with $\sigma(x)$ an arbitrary function and ${\cal R}$ an elliptic differential operator. 
In this section we take  as guiding example the operator
\be
{\cal R} = -\frac12 \pa^2 +V(x) = \frac12 p^2 + V(x)
\ee 
where $\pa^2=\pa^\mu\pa_\mu$ is the laplacian and $p_\mu= -i \pa_\mu$ the momentum operator 
in the coordinate representation of quantum mechanics.  ${\cal R}$ is directly  interpreted as the
hamiltonian of a  non-relativistic  particle of unit mass in $D$ dimensions, and the functional trace is understood 
as a trace on the Hilbert space of the particle
\be 
{\rm Tr} \left[ \sigma(x)\, e^{-\beta {\cal R}}\right] 
= \int d^D x\,  \sigma(x) \la x |e^{-\beta {\cal R}} |x\ra = \int \frac{d^D x}{(2\pi\beta)^{\frac{D}{2}}} \sigma(x)
\sum_{n=0}^{\infty}a_n(x)\beta^n
\label{5}
\ee
where $ \la x |e^{-\beta {\cal R}} |x\ra $  is the transition amplitude for an euclidean time $\beta$ 
with coinciding initial and final points, i.e. the heat kernel at coinciding points.
The evaluation of the latter for an arbitrary potential $V(x)$ is not known in closed form,
but  often one needs only its perturbative expansion for small values of $\beta$, which gives rise to 
the Seeley-DeWitt coefficients $a_n(x)$. The first few ones are
\ba
&a_0(x) = 1 \\
&a_1(x)  = - V \\ 
&a_2(x) = \frac12  V^2 -  \frac{1}{12} \partial^2   V   \\
&a_3(x) =  -\frac16  V^3 + \frac{1}{12} V\partial^2 V 
+ \frac{1}{24} \partial_\mu V \partial^\mu V 
- \frac{1}{240} \partial^4 V \;. 
\label{hkc}
\ea

A way of computing them is to use the path integral representation of the transition amplitude
for the quantum mechanical model with hamiltonian ${\cal R}$ and euclidean action 
\be 
S[x(t)] = \int_0^\beta d t  \left(\frac12 \dot{x}^\mu \dot{x}_\mu  + V(x) \right) \;.
\label{act-1}
\ee
Then, using the equivalence of path integrals with canonical quantization, one may write 
\ba 
{\rm Tr} \left[ \sigma(x)\, e^{-\beta {\cal R}}\right] 
&= \int d^D x_0\,  \sigma(x_0) 
\la x_0 |e^{-\beta {\cal R}} |x_0\ra
\cr 
&= \int d^D x_0\, \sigma(x_0) \int_{x(0) =x_0}^{x(\beta) =x_0} Dx(t)\, e^{-S[x(t)]}
= \int_{_{PBC}} Dx(t)\,  \sigma(x(0))\,  e^{-S[x(t)]}
\label{8}
\ea
where in the second line we have used the path integral representation of the transition amplitude
at coinciding points, and recognized that the additional  integration over the point $x_0$,
which creates the trace, implements periodic boundary conditions (PBC).
Thus one finds a path integral on all loops with an insertion of the function $\sigma(x(t))$. 
The argument $x^\mu(t)$ of the function $\sigma$ is evaluated at $t=0$, which corresponds to the 
base point  $x^\mu(0)=x^\mu_0$ of the parametrized loop, but it could be anywhere on the loop
described by the function
$x^\mu(t)$ as a consequence of time translational invariance.  
Now, given the relation \eqref{8}, one can use the perturbative expansion of the path integral in 
the euclidean time $\beta$  to evaluate the heat kernel coefficients in \eqref{5}
with worldline propagators and Feynman diagrams.

This set-up was discussed in \cite{Bastianelli:1991be, Bastianelli:1992ct}, where  it was extended 
to curved space and non-abelian gauge fields and used to rederive the trace anomalies of many field theories.
Actually, the precise observable available (the trace anomalies) could be used as a benchmark to construct well-defined
path integrals for particles in curved spaces, stressing the necessity of using 
precise regularization schemes on the worldline, which must include well-defined but scheme dependent
counterterms \cite{Bastianelli:1991be, Bastianelli:1992ct, deBoer:1995cb, Bastianelli:1998jm, Bastianelli:2006rx}.
A list of counterterms needed  for sigma models with $N$ supersymmetries
in various regularization  schemes  is given in \cite{Bastianelli:2011cc}, with 
the  $N=4$ case that has been applied in the recent construction of the path integral
for the graviton in first quantization \cite{Bonezzi:2018box, Bastianelli:2019xhi, Bonezzi:2020jjq}.

  A direct extension of the above construction to the case of an insertion 
of a first-order differential operator in the trace of the heat kernel may not seem immediate.
A simple way of obtaining the insertion is  to exponentiate the corresponding operator, 
 and view it as a source added to the action. Then a derivative creates the required insertion.
Let us check the formulae we get this way for a scalar insertion and compare them with
the set-up described above.

To start with,  let us consider
\be 
{\rm Tr} \left[ \sigma(x)\, e^{-\beta {\cal R}}\right]  =
 \frac{\partial}{\partial \lambda}
 {\rm Tr} \left[ e^{-\beta {\cal R} +\lambda \sigma(x)}\right] \Big |_{\lambda=0} 
\ee
where  the trace guarantees that the insertion arising by acting with a derivative can be placed on the left 
of the exponential. The exponentiation can be viewed as a deformation of the hamiltonian, 
which in turns generates a modified euclidean action with $V(x) \to V(x) - \frac{\lambda}{\beta} \sigma(x)$, so that 
\be 
S[x] \ \to \ S_\lambda [x]
= \int_0^\beta d t  \left(\frac12 \dot{x}^\mu \dot{x}_\mu  +V(x)-  \frac{\lambda}{\beta} \sigma(x) \right) \;.
\ee
Using the path integral representation one finds
 \be 
{\rm Tr} \left[ \sigma(x)\, e^{-\beta {\cal R}}\right]  =
\int_{_{PBC}}
Dx \, \left(\frac{1}{\beta} \int_0^\beta dt \  \sigma(x(t))\right)e^{-S[x]} \;.
\label{eq:11}
\ee
This formula is equivalent to the one  obtained earlier in \eqref{8}.
The equivalence between the two expressions is understood by invoking the time translational 
invariance of the one-point function of the operator $\sigma(x(t))$, which may be substituted by its time average.
 
\begin{figure}[h!]
\centering
\includegraphics{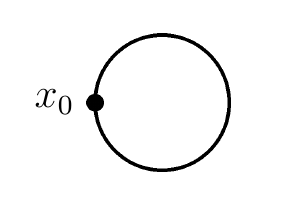}
\caption{\small Loop with Dirichlet boundary conditions (DBC) at $x_0$.}
\label{f1}
\end{figure}

As a side result, this reformulation makes it clear how to use different worldline propagators 
for obtaining the same heat kernel coefficients. In the set-up  described by eq.  \eqref{8}, it is natural to 
parametrize the quantum integration variables by
\be 
x(t) = x_0 + q(t)
\label{12}
\ee
with  $q(0)=q(\beta)=0$,  thus defining Dirichlet boundary conditions (DBC) on the quantum fluctuations
$q(t)$. They parametrize loops with a fixed base point $x_0$. 
The final integration over $x_0$ produces all possible loops in target space, thus
implementing the full periodic boundary condition (PBC) prescription, see Fig. \ref{f1}.
The emerging quantum integration variables $q(t)$ has a perturbatively well-defined propagator,
as fixed by Dirichlet boundary conditions.
This was the approach used in \cite{Bastianelli:1991be, Bastianelli:1992ct}.

\begin{figure}[h!]
\centering
\includegraphics{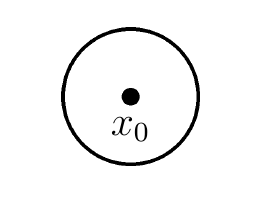}
\caption{\small Loop with average position $x_0$ (SI).}
\label{f2}
\end{figure}

Alternatively, one may find it useful to employ the so-called ``string-inspired" (SI) propagator
\cite{Schubert:2001he}, obtained  by setting again 
\be
x(t) = x_0 + q(t)
\ee
but now with the condition
\be 
x_0 = \frac{1}{\beta}
\int_0^\beta d t \, x(t) \quad \Rightarrow \quad \int_0^\beta dt \, q(t) = 0
\ee
where the zero mode $x_0$ is the average position of the loop, see Fig. \ref{f2}.
 The non-local constraint on $q(t)$ defines the 
SI propagator. 
Again, the final integration over $x_0$ creates all loops in target space.

As a preparation for our worldline calculations, we collect these propagators in appendix \ref{A},
 and use them in appendix \ref{B} for obtaining the Seeley-DeWitt  coefficients of eq. \eqref{hkc} 
 with a simple perturbative path integral calculation.

The previous set-up is easily generalized by coupling the model to curved space and 
to non-abelian gauge fields.
For the latter, the simplest strategy requires the use of a time ordering prescription to exponentiate the action with 
the matrix-valued gauge field, a  method already employed in \cite{Bastianelli:1990xn}.
More elaborate methods that avoid the time ordering are also available \cite{Bastianelli:2013pta},
and could be used as well. 
More general ways of factorizing the zero mode $x_0$ of the periodic functions $x(t)$ can be found in  
\cite{Fliegner:1997rk} and \cite{Bastianelli:2003bg}.

\section{Insertion of a first-order operator}
\label{two}

In this section, we consider the insertion of a first-order differential operator inside the trace of the heat kernel
and  construct a  path integral representation for it. 

To start with, let us consider a more general hamiltonian ${\cal R }$, with a non-abelian connection  $A_\mu$ 
and a matrix-valued scalar potential $V$
\be
{\cal R } = -\frac{1}{2}\nabla^2 +V, \hskip .6cm \nabla_\mu = \partial_\mu + A_\mu \;.
\label{15}
\ee
The corresponding  matrix-valued euclidean action for the point particle of coordinates $x^\mu(t)$ 
reads
\be 
S[x] = \int_0^{\beta}dt \left( \frac 12 \dot{x}^\mu \dot{x}_\mu + \dot{x}^\mu A_\mu (x) + V\left(x\right) \right)
\label{16}
\ee
and its exponential appears in the path integral with a time ordering. The latter guarantees gauge covariance
 as in the standard construction of Wilson lines.
The heat kernel is thus computed by the path integral on the particle coordinates  $x^\mu(t)$ as
\be
e^{-\beta {\cal R}}= \int Dx(t)  \, \T  e^{-S[x(t)]}
\ee
where T denotes the time ordering along the worldline parametrized by $t$: 
upon the expansion of the exponential one should 
place the matrices associated with earlier times on the right of those associated with later times. 
The trace of the heat kernel is computed by periodic boundary conditions with period $\beta$, 
$x^\mu(\beta)= x^\mu(0)$, and further implementing a finite dimensional trace 
(denoted by ``tr'') on the vector space where the matrix-valued potentials $A_\mu$ and $V$ act upon 
\be \label{HKT}
\Tr\left[e^{-\beta {\cal R}}\right] = \tr \int_{_{PBC}}Dx(t)  \, \T  e^{-S[x(t)]}\;. 
\ee

Next, we would like to insert on the left-hand side an operator of the form 
\be
J = \Big [\xi^\mu(x) \partial_\mu + \sigma (x)\Big]
\label{ins}
\ee
where $\xi^\mu(x) $ is an arbitrary vector field (we have in mind applications to diffeomorphism anomalies)
and  $\sigma(x)$ a matrix-valued function that we will choose appropriately to simplify the relation with the path integral
and keep gauge invariance manifest. The last contribution can be modified at will  by adding
a standard heat kernel trace with the insertion of a  matrix-valued scalar function. 
 
As in the previous section, we modify the action and the hamiltonian by adding a source
so that a derivative on its coupling constant creates an insertion. 
The source term in the action must have a coupling to $\xi^\mu(x)$, which can be considered as  
an abelian gauge field, so we deform  the action  as 
\be 
S[x] \rightarrow S _\lambda[x] =\int_0^\beta dt \left(\frac{1}{2}\dot{x}^\mu \dot{x}_\mu 
+  \dot{x}^\mu A_\mu(x) +  V(x) 
+\lambda  \dot{x}^\mu \xi_\mu(x) 
\right)
\ee 
where $\lambda$ is a coupling constant. By going through the canonical formalism, one finds that 
the hamiltonian corresponding to the previous  euclidean action is given by
\be
H = \frac12 \pi^2 + V(x)
\ee
where the covariant momentum 
\be
\pi_\mu = p_\mu  -iA_\mu(x)-i\lambda \xi_\mu(x)
\ee
becomes a covariant derivative $\overline{\nabla}_\mu$ upon quantization 
\be
\pi_\mu \quad \to  \quad 
-i \overline{\nabla}_\mu = -i  (\pa_\mu  +A_\mu(x)+ \lambda \xi_\mu(x))\;.
\ee
Fixing the ordering ambiguities to  maintain gauge covariance, one finds a corresponding 
quantum hamiltonian 
\be
{\cal R_\lambda} = -\frac{1}{2} \overline{\nabla}^2 +V  = 
{\cal R} -\lambda\left(\xi^\mu(x) \nabla_\mu + \frac 12(\partial_\mu \xi^\mu(x)) \right) -\frac {\lambda^2}{ 2 } \xi^2(x) 
\ee
and a deformed version of \eqref{HKT} may be written down 
\be \label{LHKT} 
\Tr\left[ e^{-\beta {\cal R_\lambda}} \right] = \tr \int_{_{PBC}}Dx  \, \T e^{-S_\lambda[x]}\;.
\ee
Taking a $\lambda$-derivative on both sides, and setting $\lambda=0$, 
one finds on the left-hand side the insertion of the operator
\be
\xi^\mu \nabla_\mu + \frac 12(\partial_\mu \xi^\mu) 
\ee
and on the right-hand side its path integral realization
\be 
\Tr\left[\left(\xi^\mu \nabla_\mu	+ \frac12(\partial_\mu \xi^\mu) \right) e^{-\beta {\cal R}}\right] 
=
\tr \int_{_{PBC}}Dx \left(- \frac{1}{\beta} \int_0^\beta dt \, \dot{x}^\mu \xi_\mu(x) \right) \T e^{-S[x]}
\ee
which is the formula we were looking for.

The insertion on the path integral side may be simplified
by  using time translation invariance on the worldline, and one may substitute 
\be 
- \frac{1}{\beta} \int_0^\beta dt \, \frac{d x^\mu (t)}{dt} \xi_\mu(x(t)) 
 \qquad \to \qquad 
-\frac{d x^\mu (0)}{dt} \xi_\mu(x(0)) 
\label{28}
  \ee
  with the insertion evaluated at a point of the loop, chosen here as the initial point.
    In the DBC method for evaluating the path integral one can use   
  $x^\mu(0) = x^\mu_0$, see Fig. \ref{f1}. 
In the SI method one will have to set  $x^\mu(0) = x^\mu_0 + q^\mu(0)$ instead,  see Fig. \ref{f2}.

To summarize, we have found that computing with the path integral the expectation value of 
\be
- \frac{1}{\beta} \int_0^\beta dt \, \dot{x}^\mu \xi_\mu(x)   \qquad {\rm or} \qquad
 -\frac{d x^\mu (0)}{dt} \xi_\mu(x(0)) 
\ee
creates an insertion of the operator
\eqref{ins} with the matrix-valued function $\sigma(x)$ fixed to be
\be
 \sigma(x) = \xi^\mu A_\mu +\frac12 (\pa_\mu \xi^\mu) \;.
\ee
We will study generalized heat kernel coefficients corresponding to this particular insertion.
Other forms of $\sigma(x)$ can be easily worked out. 

\section{Generalized heat kernel coefficients}
\label{three}

Having found a path integral representation of the trace of the heat kernel with the insertion 
of a first-order differential operator, we evaluate the corresponding heat kernel coefficients
by using the perturbative expansion  in $\beta $ of the path integral. It takes the form 
\be 
\Tr\left[\left( \xi^\mu \nabla_\mu	+ \frac{1}{2}(\partial_\mu \xi^\mu) \right) 
e^{-\beta {\cal R}}\right] = \int \frac{d^Dx }{\left(2\pi\beta\right)^{\frac{D}{2}}} \sum_{n=0}^{\infty} 
b_n (x) \beta^n  
\ee
where the $b_n(x) $ are the generalized heat kernel coefficients which include at the linear order  
the abelian vector field $\xi^\mu$.  For the operator $\cal R $ in \eqref{15} we use the action 
in \eqref{16} and compute up to order $\beta^3$ to find
\begin{align} 
\label{ghkc}
b_{0} &=  0 \nonumber\\
b_{1} &=   
\frac{1}{24}\, \tr\, 
G^{\mu\nu}F_{\mu\nu} \nonumber\\
b_{2} &=  \frac{1}{480}\, \tr\, \Big [ 
\partial^2 G^{\mu\nu} F_{\mu\nu} - 20 G^{\mu\nu}  F_{\mu\nu}  V  
 \Big ] \\
 b_{3} &= \frac{1}{1440}\,
\tr\,  \Big[
\frac{3}{28}\partial^4G^{\mu\nu} F_{\mu\nu} 
+ \frac{5}{4}G^{\mu\nu}F_{\mu\nu}F^2 
+ G^{\mu\nu}F_{\nu\rho}F^{\rho\lambda}F_{\lambda\mu}
+30G^{\mu\nu}F_{\mu\nu}V^2
 \nonumber \\
&- \Big ( 6G^{\mu\nu}  \nabla^2 F_{\mu\nu} 
+6 \partial^2 G^{\mu\nu}  F_{\mu\nu}
+8\partial^\lambda G^{\mu\nu}\nabla_\lambda F_{\mu\nu}  
 +2\partial_\mu G^{\mu\nu}\nabla^\lambda F_{\lambda\nu} 
 +6G^{\mu\nu}F_{\mu\lambda}F ^\lambda{}_\nu \Big )V 
\Big]
\nonumber 
\end{align}
where $G_{\mu\nu} = \pa_\mu \xi_\nu -\pa_\nu \xi_\mu$
is the abelian field strength, 
$F_{\mu\nu} = \pa_\mu A_\nu -\pa_\nu A_\mu + [A_\mu, A_\nu] $
the non-abelian field strength,
and $\nabla_\mu$ the covariant derivative of  $A_\mu$. 
Of course,  $G_{\mu\nu}$ could be taken out of the color trace ``tr". 
These coefficients are up to total derivatives, and we have freed $V$ from derivatives.

The coefficient $b_1$ was given in \cite{Bastianelli:1990xn} and
$b_2$ in \cite{Branson:1997ze}, both including their coupling to gravity.
The coefficient $b_3$ is new, as far as we know. 
In the next sections, we describe their explicit evaluation through the perturbative expansion 
of the path integral.

We have used an abelian vector field $\xi^\mu$, which allows for simplifications in the above formulae. 
For example, in $b_1$  the term $G_{\mu\nu}$ can be taken out of the color trace so that only the abelian part 
of $F_{\mu\nu}$ survives the trace. Similarly, one may simplify the other coefficients, or write them in equivalent ways.
These coefficients may also be generalized by considering a non-abelian vector field $\xi^\mu$, 
as in  \cite{Branson:1997ze}, but we have chosen to keep it abelian for a direct
application to the anomalies in the conservation of the stress tensor.

\section{Perturbative expansion} 
\label{four}

We now study the perturbative expansion of the path integral with PBC, i.e. considering worldlines with 
the topology of a circle. Since the kinetic operator cannot be inverted on the circle, 
one has to factor out a zero mode $x_0^\mu$ and split the path integration variable as
\be 
x^\mu(t) = x^\mu_0 + q^\mu(t) \;.
\ee
This can be done using either the DBC method or the SI one, as explained earlier.

To start with we rescale the time  $t\to \tau =\frac{t}{\beta}$, so that  $\tau\in [0,1]$, and  we find 
the following path integral representation of the trace
\ba
\Tr\left[\left( \xi^\mu \nabla_\mu	+ \frac{1}{2}(\partial_\mu \xi^\mu) \right) 
e^{-\beta {\cal R}}\right] 
&= 
\int d^Dx_0 \
\tr 
\int_{_{DBC}} Dq 		
\left(-\frac{1}{\beta}\int_0^1 d\tau \, \dot{q}^\mu \xi_\mu(x_0+q) \right)	
\T 
e^{-S[x_0+q]}
\\
 &= \int \frac{d^Dx_0 }{\left(2\pi\beta\right)^{\frac{D}{2}}}\,
 \tr  
 \left
 \la \left(-\frac{1}{\beta}\int_0^1 d\tau \, \dot{q}^\mu \xi_\mu(x_0+q) \right)	\T 
e^{-S_{int}[x_0+q]}
\right \ra
\label{33}
\ea
where dots indicate derivative with respect to $\tau$, and angle brackets denote 
normalized averages with the free action, $\la1\ra=1$. 
The expectation values are to be computed by Wick contracting with the propagators in appendix \ref{A},
and with the interaction vertex taking the form 
\be 
S_{int}[x_0+q] =  \int_0^1 d\tau 
\Big (  \dot{q}^\mu A_\mu (x_0+q) + \beta V (x_0+q) \Big ) \;.
\ee

We now start computing at order $\beta$ to get $b_1$. 
There are two contributions. The first one comes from Taylor expanding  
$\xi_\mu$  and $A_\mu$ to first order in $q^\mu$, and gives for the right-hand side of \eqref{33} 
  \be
  \int \frac{d^Dx_0 }{\left(2\pi\beta\right)^{\frac{D}{2}}}\,
 \tr  
 \left
 \la 
 \left(-\frac{1}{\beta}  \pa_\nu \xi_\mu(x_0)  \int_0^1 d\tau \, q^\nu(\tau)  \dot{q}^\mu(\tau) 
 \right)	
 \left(- \pa_\beta A_\alpha(x_0) \int_0^1 d\tau' \,  q^\beta(\tau') \dot{q}^\alpha(\tau') 
  \right) \right \ra \;.
  \label{36}
  \ee
Time ordering is not needed and the disconnected Wick contractions vanish 
\be
{\bf F}_1= \fone =  \int_0^1 d\tau\, \deld (\tau,\tau) = 0   \;.
\ee
The remaining connected correlation function gives
\be
 \la    q^\nu(\tau)  \dot{q}^\mu(\tau) q^\beta(\tau')  \dot{q}^\alpha(\tau')  \ra_c
 =\beta^2 
 \Big ( \delta^{\nu \beta}  \delta^{\mu \alpha} \,
 \del(\tau,\tau') \,  \ddeld(\tau,\tau')  
 +
 \delta^{\nu \alpha} \delta^{\mu \beta} \, 
\deld(\tau,\tau') \, \ddel(\tau,\tau')   \Big )  
 \ee
 where  the first term corresponds to a worldline Feynman diagram of the form
\be
{\bf F}_2= \ftwo = \int_0^1 d\tau\, \int_0^1 d\tau'\, \del(\tau,\tau')\, \ddeld(\tau,\tau')  
 = \frac{1}{12}
\ee
and the second one to a diagram of the form
 \be
{\bf F}_3= \fthree =\int_0^1 d\tau\, \int_0^1 d\tau'\,  \deld(\tau,\tau')  \,\ddel(\tau,\tau') 
=  -\frac{1}{12} \;.
\label{f3}
\ee
In drawing  Feynman diagram we denote vertices by black dots and derivatives by white circles on the legs.
Integration by  parts relates the two integrals, ${\bf F}_2=-{\bf F}_3$, hinting at gauge invariance. 
The above values are obtained using 
equivalently the DBC or the SI propagators. In the latter case one may use translational invariance
to eliminate one integration.  Thus, the trace inside \eqref{36} reduces to 
\be 
\beta\, \tr \,\partial^\nu \xi^\mu(x_0) \Big (\partial_\nu A_\mu(x_0) {\bf F}_2 +
\partial_\mu A_\nu(x_0) {\bf F}_3\Big )
=
\frac{\beta}{12}\tr \,\partial^\nu \xi^\mu(x_0) \Big (\partial_\nu A_\mu(x_0) - \partial_\mu A_\nu(x_0) \Big )\;.
\label{b1-1}
\ee

A second term  of the same order in $\beta$ arises from considering two interaction vertices and has the effect 
of completing the non-abelian gauge invariance. 
 Keeping the leading term of the Taylor expansion  of the non-abelian potential inside $\frac12 S_{int}^2[x_0+q]$
 one finds 
\be
\frac{1}{2}\, \textrm{T}\!
\int_0^1 d\tau_1
 \dot{q}^\alpha (\tau_1) A_\alpha(x_0)
 \int_0^1 d\tau_2 \, \dot{q}^\beta(\tau_2)
A_\beta(x_0) \;.
\ee
The time ordering is implemented explicitly as 
\be
\frac{1}{2} A_\alpha(x_0) A_\beta(x_0)
\int_0^1 d\tau_1 \int_0^{\tau_1} \! d\tau_2 \ \dot{q}^\alpha (\tau_1)  \dot{q}^\beta(\tau_2)
+
\frac{1}{2} A_\beta(x_0) A_\alpha(x_0)
\int_0^1 d\tau_1 \int_{\tau_1}^1 \! d\tau_2 \ \dot{q}^\alpha (\tau_1)  \dot{q}^\beta(\tau_2)
\ee
and simplified using a Heaviside step function (and renaming integration variables) 
\be 
A_\alpha(x_0)A_\beta(x_0) 
\int_0^1 d\tau_1 \int_0^1 d\tau_2\ 
\dot{q}^\alpha(\tau_1)\dot{q}^\beta(\tau_2)  \theta(\tau_1-\tau_2) \;.
\ee
Inserted into the right-hand side of \eqref{33} it leads to 
\be 
\frac{\left(-1\right)}{\beta}\tr \,\partial_\nu \xi_\mu(x_0)A_\alpha(x_0)A_\beta(x_0)\int_0^1 d\tau \int_0^1 d\tau_1 \int_0^1 d\tau_2 \,
\la \dot{q}^\mu(\tau)q^\nu(\tau)\dot{q}^\alpha(\tau_1)\dot{q}^\beta(\tau_2) \ra\, \theta(\tau_1-\tau_2) 
\ee
with nonvanishing contractions that produce the following integrals 
\ba 
\int_0^1 d\tau \int_0^1 d\tau_1 \int_0^1 d\tau_2\ \ddeld(\tau,\tau_1)\deld(\tau,\tau_2)\theta(\tau_1-\tau_2) &=\frac{1}{12}\\
\int_0^1 d\tau \int_0^1 d\tau_1 \int_0^1 d\tau_2\ \ddeld(\tau,\tau_2)\deld(\tau,\tau_1)\theta(\tau_1-\tau_2) &=-\frac{1}{12}
\ea
independently of the propagator used. At the end one finds a commutator term 
\be 
\frac{\beta}{12}\tr \,\partial^\nu \xi^\mu(x_0) [A_\nu(x_0), A_\mu(x_0)] \;.
\label{comm}
\ee

These are all the terms of order $\beta$. Summing \eqref{b1-1} and \eqref{comm}
   one finds for \eqref{33}
 \be
  \int\frac{d^Dx_0 }{\left(2\pi\beta\right)^{\frac{D}{2}}}\,
 \tr \,  
 \frac{\beta}{12} \partial^\nu\xi^\mu(x_0) F_{\nu\mu}(x_0) =\int\frac{d^Dx_0 }{\left(2\pi\beta\right)^{\frac{D}{2}}}\,
 \tr \,  
 \frac{\beta}{24} G^{\mu\nu}(x_0) F_{\mu\nu}(x_0)
 \ee
 which delivers  the coefficient $b_1$  of eq. \eqref{ghkc}.
 $G_{\mu\nu}$ is the abelian field strength of $\xi_\mu$ and
can be taken out of the trace, showing that only the abelian part of $F_{\mu\nu}$
contributes. The time ordered diagram that leads to the commutator
in \eqref{comm} does not survive the trace, but we have presented it to exemplify the role of the  
time ordering prescription.

As noted, the SI propagators are explicitly translational invariant and in the perturbative expansion one may  
eliminate an integration of the Feynman diagrams, 
fixing for example the insertion at $\tau=0$. 
In the DBC method, translational invariance on the circle can be used as well.
However, the calculation proceeds somewhat differently. One uses translational invariance
to fix the insertion at $\tau =0$ and identifies $x^\mu(0) = x^\mu_0$ (since $q^\mu(0) = 0$ by DBC). 
 Then, eq. \eqref{33} simplifies to
\ba
\Tr\left[\left( \xi^\mu \nabla_\mu	+ \frac{1}{2}(\partial_\mu \xi^\mu) \right) 
e^{-\beta {\cal R}}\right] 
 = \int \frac{d^Dx_0 }  {\left(2\pi\beta\right)^{\frac{D}{2}}}\,
 \xi_\mu(x_0)
 \frac{(-1)}{\beta}  
 \tr   \left \la 
 \dot{q}^\mu(0)  
 \T  e^{-S_{int}[x_0+q]}
\right \ra_{_{\!\! DBC}}
\label{pi-42}
\ea
which delivers the result with the vector field $\xi_\mu(x_0)$ explicitly factored out.
To evaluate the same coefficient $b_1$ in this set-up,
 one needs to expand $A_\mu$ to higher orders.
The calculation is simplified  by using the Fock-Schwinger gauge (more on this later)
\be 
A_\nu(x_0+q) = A_\nu(x_0) + \frac 12 q^\rho F_{\rho\nu}(x_0) +\frac 13 q^\rho q^\sigma
 \nabla_\sigma F_{\rho\nu}(x_0)
+...
\ee
 to find
\ba
\frac{(-1)}{\beta}   \left \la \dot{q}^\mu(0) \T  e^{-S_{int}[x_0+q]} \right \ra_{_{\!\! DBC}} 
&= 
\frac{1}{\beta}   \left \la \dot{q}^\mu(0) S_{int}[x_0+q] +\cdots\right \ra_{_{DBC}} \\
&= 
 \frac{1}{3\beta} \nabla_\sigma F_{\rho\nu}(x_0)\int_0^1 d\tau  \,
\la \dot{q}^\mu(0)  \dot{q}^\nu(\tau) q^\rho(\tau) q^\sigma(\tau)
 \ra_{_{DBC}} \\
 &= \frac{\beta}{3} \nabla_\nu F^{\nu \mu}(x_0) ({\bf G}_1 -{\bf G}_2)
= -\frac{\beta}{12} \nabla_\nu F^{\nu \mu}(x_0) 
 \ea
where the integrals with the DBC propagators give 
\ba
{\bf G}_1 &=\gone= \int_0^1 d\tau\, \ddeld(0,\tau)\del(\tau,\tau) =-\frac 16\\
{\bf G}_2 &= \gtwo=\int_0^1 d\tau\, \ddel(0,\tau)\deld(\tau,\tau) =\frac{1}{12}
\ea
(the cross in the vertex singles out  the vertex without integration). 
Considering the color trace in \eqref{pi-42}
one may substitute the covariant derivative with the standard derivative
and notice that  only the abelian part of $ F^{\mu \nu}$  survives the trace, so that by setting 
$ b_{1} = b_{1\mu}\, \xi^\mu $  
one finds  
\be
b_{1\mu}= -\frac{1}{12} \, \tr\, \pa^\nu F_{\nu \mu} 
\ee
that indeed reproduces  $b_1$ after integrating by part inside \eqref{pi-42}.

The coefficient $b_1$ was known from ref. \cite{Bastianelli:1990xn}, 
where it was obtained by computing the heat kernel trace with plane waves, but presented in a form that did not
show manifest  gauge invariance.
The previous calculation verifies the consistency of the path integral method, 
and one may proceed with confidence to evaluate higher order coefficients.

\section{Higher order coefficients}
\label{five}

To get additional coefficients one needs to push the perturbative expansion  in $\beta$ to higher orders. 
To proceed faster,  we use  gauge invariance and select the Fock-Schwinger (FS) gauge 
\be 
q^\mu(\tau)A_\mu(x_0 + q(\tau)) = 0 
\ee
which allows to expand the gauge potential in terms of its curvature (and derivatives 
thereof) evaluated at $x_0$ (see for example \cite{Muller:1997zk})
\be 
A_\mu(x_0+q) = A_\mu(x_0) + \frac 12 q^\nu F_{\nu\mu}(x_0) +\frac 13 q^\nu q^\rho \nabla_\rho F_{\nu\mu}(x_0)
+\frac{1}{8} q^\nu q^\rho q^\sigma \nabla_\sigma\nabla_\rho F_{\nu\mu}(x_0) +... \;.
\ee
A similar gauge  holds also for $\xi_\mu (x)$
\be 
\xi_\mu(x_0+q) = \xi_\mu(x_0) + \frac 12 q^\nu G_{\nu\mu}(x_0) +\frac 13 q^\nu q^\rho \pa_\rho G_{\nu\mu}(x_0)
+\frac{1}{8} q^\nu q^\rho q^\sigma \pa_\sigma \pa_\rho G_{\nu\mu}(x_0) +...\;.
\ee
Next, we give some details on the calculation of the higher order coefficients $b_2$ and $b_3$.

To identify  $b_2$ we need terms with one vertex (from $S_{int}$)  and two vertices (from $S^2_{int}$).
Substituting the potentials in the FS gauge, both in the insertion and in the vertices, we get
the following contribution
of order  $\beta^2$ to eq. \eqref{33}
\be
 \int \frac{d^Dx_0 }{\left(2\pi\beta\right)^{\frac{D}{2}}}\,
 \tr \Big (A_1(x_0) +A_2(x_0) +A_3(x_0) +A_4(x_0) \Big)
 \label{56}
\ee
where the single vertex produces\footnote{We now use a compact notation,
indicating $q_0\equiv q(\tau_0)$, $\int_{01} \equiv \int_0^1 d\tau_0 \int_0^1 d\tau_1 $, etc..} 
\be \label{a1}
A_1(x_0) 
= \frac{1}{16\beta} \int_{01}  
\la \dot{q}^\mu_0 q^\nu_0 \dot{q}^\alpha_1 q^\beta_1 q^\gamma_1 q^\delta_1 \ra 
G_{\nu\mu}(x_0)  \nabla_\delta \nabla_\gamma F_{\beta \alpha}(x_0) \ee
\be \label{a2}
A_2(x_0) =\frac{1}{9\beta}\int_{01}
\la \dot{q}^\mu_0 q^\nu_0 q^\rho_0 \dot{q}^\alpha_1 q^\beta_1 q^\gamma_1\ra 
\partial_\rho G_{\nu\mu}(x_0) \nabla_\gamma F_{\beta\alpha}(x_0)\ee
\be \label{a3}
A_3(x_0)= \frac{1}{16\beta}\int_{01}\la \dot{q}^\mu_0 q^\nu_0 q^\rho_0 q^\sigma_0  \dot{q}^\alpha_1  q^\beta_1 \ra 
\partial_\sigma\partial_\rho G_{\nu\mu}(x_0)F_{\beta\alpha}(x_0)\ee
while the two interaction vertices contribute with 
\be \label{a4}
A_4(x_0)=-\frac{1}{4}\int_{01}\la \dot{q}^\mu_0 q^\nu_0 \dot{q}^\alpha_1 q^\beta_1\ra G_{\nu\mu}(x_0)
F_{\beta\alpha}(x_0)V(x_0) \;.
\ee
In this last term we have used cyclicity of the trace in \eqref{56} to eliminate the time ordering. Then, it
describes a disconnected contribution that embeds $b_1$, so it is immediately evaluated to 
\be
A_4 = -\frac{\beta^2}{24}G^{\mu\nu}F_{\mu\nu}V \;.
\ee
As for the remaining terms, since they enter eq.\eqref{56},
we simplify them with integration by parts (covariant derivatives acting on $F_{\mu\nu}$
become usual derivatives acting on $G_{\mu\nu}$). Then, renaming the time variables we get
\be 
A_1 + A_3
= \frac{1}{16\beta} \int_{01}  
\la \dot{q}^\mu_0 q^\nu_0 \dot{q}^\alpha_1 q^\beta_1 q^\gamma_1 q^\delta_1 \ra
\Big (\pa_\gamma \pa_\delta G_{\nu\mu}  F_{\beta \alpha} + \pa_\gamma \pa_\delta G_{\beta\alpha} F_{\nu\mu} \Big )
\ee
and performing the Wick contractions (and also integrating by parts on the worldline to get rid of $\ddeld_{01}$
in the Feynman diagrams)
we get
\be 
A_1 + A_3
= \frac{\beta^2}{4} {\bf H}_1 (\pa^2 G^{\mu\nu}  F_{\mu\nu} + 2 \pa^\alpha\pa_\beta G^{\beta\nu}  F_{\alpha\nu})
=  \frac{\beta^2}{2} {\bf H}_1\, \pa^2 G^{\mu\nu}  F_{\mu\nu} 
\ee
where ${\bf H}_1$ is the Feynman diagram 
\be
{\bf H}_1 
=\jone  
=\int_{01} \ddel_{01}\deld_{01} \del_{11}  
= 
\begin{cases}
\frac{1}{60} & DBC\\
\frac{1}{144} & SI \;.
\end{cases}
\ee
We  proceed similarly with $A_2$ to get 
\ba 
A_2 &=- \frac{1}{9\beta}\int_{01}
\la \dot{q}^\mu_0 q^\nu_0 q^\rho_0 \dot{q}^\alpha_1 q^\beta_1 q^\gamma_1\ra 
\pa_\gamma \partial_\rho G_{\nu\mu}
F_{\beta\alpha}
\\
&=
\frac{\beta^2}{9} \pa^2 G^{\mu\nu} F_{\mu\nu}  \Big (-\frac{9}{2} {\bf H}_2 
+\frac12 {\bf H}_3 -  \frac14{\bf H}_4 +2 {\bf H}_5\Big)
\ea
where
\ba
&{\bf H}_2 =\jtwo=\int_{01} \ddel_{01} \del_{01} \deld_{01}
=
\begin{cases}
\frac{1}{90} & DBC\\
\frac{1}{360} & SI
\end{cases}
\\[2mm]
&
{\bf H}_3 =\jthree=\int_{01} \del_{00} \ddeld_{01} \del_{11} =
\begin{cases}
-\frac{1}{180} & DBC\\
0 & SI
\end{cases} 
\\[2mm]
&
{\bf H}_4 =\jfour=\int_{01} \del_{00} \ddel_{01} \deld_{11} =
\begin{cases}
\frac{1}{360} & DBC\\
0 & SI
\end{cases}
\\[2mm]
&
{\bf H}_5 =\jfive=\int_{01} \ddel_{00} \del_{01} \deld_{11} 
=
\begin{cases}
-\frac{1}{720} & DBC\\
0 & SI \;.
\end{cases}
\ea
Adding all terms up we  get
$$A_1+A_2+A_3 = 
\beta^2 \pa^2 G^{\mu\nu} F_{\mu\nu}  \Big (\frac{1}{2} {\bf H}_1 -\frac{1}{2} {\bf H}_2 
+\frac{1}{18} {\bf H}_3 - \frac{1}{36} {\bf H}_4  +\frac{2}{9} {\bf H}_5\Big)  
=
\frac{\beta^2}{480} 
\partial^2 G^{\mu\nu} F_{\mu\nu} 
$$
independently of the worldline propagators used. 
Including $A_4$ we obtain the generalized coefficient $b_2$
\be
 b_{2} =  \frac{1}{480}\,\tr\,  \Big [ 
\partial^2 G^{\mu\nu} F_{\mu\nu} - 20 G^{\mu\nu}  F_{\mu\nu}  V  
 \Big ] 
 \ee
which correctly reproduces the one reported in \cite{Branson:1997ze} 
(with abelian $\xi^\mu$ and in flat space).

Finally, we wish to compute the coefficient $b_3$, which did not appear in the literature so far. It is 
 more laborious, so we just present the calculation of a single term, i.e. the first one 
inside $b_3$ of eq. \eqref{ghkc}, dumping details on the calculation of the remaining part into appendix \ref{C}.
This term receives contributions from the  $F_{\mu\nu}$ dependence of a   
single $S_{int}$ vertex  insertion, which read

\be\label{B1}
B_1(x_0) =\frac{1}{288\beta}\int_{01}
\la \dot{q}^\mu_0 q^\nu_0 
\dot{q}^\alpha_1 q^\beta_1 q^\gamma_1 q^\delta_1 q^\epsilon_1 q^\eta_1\ra \,
G_{\nu\mu}
\nabla_\eta  \nabla_\epsilon \nabla_\delta \nabla_\gamma F_{\beta\alpha}
\ee

\be \label{B3}
B_2(x_0)=\frac{1}{90\beta}\int_{01}\la \dot{q}^\mu_0 q^\nu_0 q^\rho_0 
\dot{q}^\alpha_1 q^\beta_1 q^\gamma_1 q^\delta_1 q^\epsilon_1\ra \,
\partial_\rho G_{\nu\mu}
\nabla_\epsilon\nabla_\delta\nabla_\gamma F_{\beta\alpha}
\ee

\be \label{B4}
B_3(x_0)=\frac{1}{64\beta}\int_{01}
\la \dot{q}^\mu_0 q^\nu_0 q^\rho_0 q^\sigma_0 \dot{q}^\alpha_1 q^\beta_1 q^\gamma_1 q^\delta_1 \ra \,
 \partial_\sigma\partial_\rho G_{\nu\mu}
 \nabla_\delta\nabla_\gamma F_{\beta\alpha}
\ee

\be \label{B5} 
B_4(x_0)=\frac{1}{90\beta}\int_{01}\la \dot{q}^\mu_0 q^\nu_0 q^\rho_0 q^\sigma_0 q^\tau_0 
\dot{q}^\alpha_1 q^\beta_1 q^\gamma_1\ra \,
\partial_\tau\partial_\sigma \pa_\rho G_{\nu\mu}\nabla_\gamma F_{\beta\alpha}
\ee

\be \label{B2}
B_5(x_0)=\frac{1}{288\beta}\int_{01}\la 
\dot{q}^\mu_0 q^\nu_0 q^\rho_0 q^\sigma_0  q^\tau_0 q^\lambda_0 
\dot{q}^\alpha_1 q^\beta_1\ra \,
\partial_\lambda \partial_\tau \partial_\sigma \partial_\rho G_{\nu\mu}F_{\beta\alpha} \;.
\ee
As they are integrated in spacetime, see eq. \eqref{33}, we integrate by parts the 
covariant derivatives from $F_{\beta\alpha}$ to $G_{\nu\mu}$, where they become standard derivatives. 
Then collecting identical Wick contractions we get  
\ba 
B_1 + B_5 &=-\frac{\beta^3}{8}\partial^4 G^{\mu\nu} F_{\mu\nu} {\bf I}_1\\
B_2 +B_4 &= -\frac{\beta^3}{30}\partial^4 G^{\mu\nu}F_{\mu\nu}\left(-15{\bf I}_2-3{\bf I}_3	+{\bf I}_4 -{\bf I}_5\right)\\
B_3 &=-\frac{\beta^3}{8}\partial^4 G^{\mu\nu}F_{\mu\nu}\left(2{\bf I}_{6}	 + {\bf I}_{7} +2{\bf I}_{8}	\right)
\ea
where the integrals corresponding to the worldline Feynman diagrams are listed in appendix \ref{C}.
Adding these terms together we get the following contribution to $b_3$
\be 
\frac{1}{1440}\int \frac{d^Dx}{\left(2\pi\beta\right)^{\frac{D}{2}}}\tr \frac{3}{28}\partial^4 G^{\mu\nu}F_{\mu\nu} \;.
\ee
In appendix \ref{C} we report details on the calculation of the other terms contributing to $b_3$.

\section{Conclusions}

We have studied path integral methods to compute heat kernel traces with insertion of a first-order differential operator.
We have considered hamiltonians ${\cal R}$ with couplings to non-abelian gauge fields and matrix-valued potentials only.
The coupling to a curved space metric is  however straightforward, as path integrals on curved spaces are 
well-studied by now \cite{Bastianelli:2006rx}.
The insertion of a first-order differential operator into the trace of the heat kernel has been 
obtained by modifying the hamiltonian with a source coupled to the first-order  operator, and then 
varying the source. This procedure translates then into the path integral representation
of the desired trace, which we have used
to calculate  the first three generalized heat kernel coefficients. 
Alternatively,  one could have applied the variational procedure directly to the standard heat kernel coefficients,
as the source term is structurally similar to the gauge coupling already present in the hamiltonian. Indeed, this 
was the method followed in  \cite{Branson:1997ze}. In the present case our results can be checked, 
and further extended to reach $b_4$ and $b_5$, by  a gauge variation of the coefficients already calculated 
with worldline methods in \cite{Fliegner:1997rk}.
We have verified the correctness of our calculation this way as well.
 
Our interest in these particular traces stems from a desire to compute the anomalies in the conservation of the 
stress tensor, which appear in four dimensions if one uses regularization schemes  that are not symmetric enough. 
Such a situation emerged in the study of a Weyl fermion in a U(1) background once regulated with Pauli-Villars fields 
with Dirac mass  \cite{Bastianelli:2018osv}. The gravitational anomaly emerging in this scheme
was calculated using generalized heat kernel coefficients in \cite{Montefiori:2019}.
The study of the anomaly structure of chiral fermions in four dimensions has become recently of renewed interest, 
in particular regarding the trace anomaly. The latter has been scrutinized  from various perspectives 
\cite{Bastianelli:2016nuf, Bastianelli:2018osv, Godazgar:2018boc, Frob:2019dgf, Bastianelli:2019fot, Bastianelli:2019zrq}
to verify the absence of the Pontryagin topological density  (in curved space) or 
Chern-Pontryagin topological density (for couplings to gauge fields).
The presence of these topological densities was
 conjectured to  be a possibility in \cite{Nakayama:2012gu}
 (see also \cite{Nakagawa:2020gqc} for a supersymmetric extension of the conjecture), and
 the analyses of refs. \cite{Bonora:2014qla, Bonora:2017gzz, Bonora:2018obr} claimed  
 their existence in  the trace anomaly of a Weyl fermion in curved space. 
 It seems useful to consider these issues even within regularization schemes that induce anomalies in the conservation 
 of the stress tensor.

The methods presented here may be considered as part of a general strategy of using 
worldline path integrals to obtain field theoretical results in flat \cite{Schubert:2001he} 
and curved space \cite{Bastianelli:2002fv}, a strategy often referred to as the worldline formalism.
These methods are quite efficient from a calculational point of view, and it seems worthwhile 
to extend their development and applications.

\vfill\eject

\appendix
\section{Worldline propagators}\label{A}

For perturbative computations in $\beta$ we find it useful to rescale the time  $t\to \tau =\frac{t}{\beta}$, 
so that  $\tau\in [0,1]$. Then,  $\beta $ appear explicitly as a perturbative parameter multiplying suitably 
the various terms of the action \eqref{act-1}, which takes the form 
\be 
S[x(\tau)]  = \int_0^1 d \tau  \left(\frac{1}{2\beta} \dot{x}^\mu \dot{x}_\mu  + \beta V(x) \right) 
\label{act-2}
\ee
where now  $\dot{x}^\mu \equiv \frac{d x^\mu}{d\tau}$. We consider periodic boundary conditions
for $x^\mu(\tau)$, appropriate for creating a functional trace in the path integral.   
The kinetic term identifies the propagator, which then carries a power of $\beta$,
while the potential term is treated perturbatively.
Setting 
\be
x^\mu(\tau) = x^\mu_0 + q^\mu(\tau)
\ee
one finds a perturbative propagator for the quantum field $q^\mu(\tau)$
of the form 
\be 
\la q^\mu(\tau)q^\nu(\tau')\ra = -\beta \delta^{\mu\nu} \del(\tau,\tau')
\ee
where $\del(\tau,\tau')$ is the Green function of the operator  $\frac{d^2}{d\tau^2}$ 
that depends on the boundary conditions and the way the zero mode $x^\mu_0$ is factored out
(recall that the differential operator $\frac{d^2}{d\tau^2}$ is not invertible on the space of periodic functions, 
the constant function has zero eigenvalue and constitutes a zero mode  of the operator). 
\\

{\em Dirichlet boundary conditions}\\[1mm]  
Using the Dirichlet boundary conditions (DBC), $q^\mu(0)=q^\mu(1)=0$, one finds for $\del(\tau,\tau')$ 
\ba 
\del_{D}(\tau,\tau') &=(\tau-1)\tau'\,  \theta(\tau-\tau') + (\tau'-1)\tau \, \theta(\tau'-\tau)
\\
&=\frac12 |\tau-\tau'| - \frac12 (\tau+\tau') +\tau\tau'
\ea
with the step function $\theta(\tau)$
 defined such that $\theta(0) = \frac12$. It satisfies
\be
\frac{d^2}{d\tau^2}  \del_{D}(\tau,\tau') = \delta(\tau-\tau')
\ee
where the Dirac delta is the one appropriate for functions with vanishing boundary conditions. 
For later use it is convenient to list the derivatives of the  worldline propagator in DBC,
where a left/right dot indicates a derivative with respect to the first/second argument
\be \begin{aligned} \label{A.11}
\ddel_D(\tau,\tau') &= \tau' -\theta(\tau'-\tau)\\
\deld_D(\tau,\tau') &=\tau - \theta(\tau-\tau')\\
\ddeld_D(\tau,\tau') &= 1 - \delta(\tau-\tau')\\
\dddel_D(\tau,\tau') &= \delta(\tau-\tau')
\end{aligned}\ee
with coincident points limits
\be\begin{aligned}
\del_D(\tau,\tau) &= \tau^2 - \tau\\
\ddel_D(\tau,\tau)&=  \tau - \frac{1}{2}\;.
\end{aligned}\ee

{\em String inspired propagator}\\[1mm]  
The ``string inspired" (SI) propagator for the quantum fluctuations $q^\mu(\tau)$
satisfies periodic boundary conditions but with the constraint
$\int_0^1 d\tau \, q^\mu(\tau) = 0$.  Then, one finds for $\del(\tau,\tau')$ 
\be 
\del_{SI}(\tau,\tau') = 
\del_{SI}(\tau-\tau') = \frac12 |\tau-\tau'| - \frac12 (\tau-\tau')^2 -\frac{1}{12}
\ee
which satisfies
\be
\frac{d^2}{d\tau^2}  \del_{SI}(\tau-\tau') = \delta(\tau-\tau') -1 \;.
\ee
It has the useful  property of being translational invariant.
It is an even function of $\tau-\tau'$, and its first derivative is odd which implies that its coincident points limit vanishes.
Here is a list of its properties 
\ba
\ddel_{SI}(\tau-\tau') &= \frac12 {\rm sgn}(\tau-\tau') - (\tau-\tau') \\
\dddel_{SI}(\tau-\tau') &=  \delta(\tau-\tau') -1 \\
\del_{SI}(0) &= -\frac{1}{12}\\
\ddel_{SI}(0) &= 0 
\ea
where by ${\rm sgn}(x)$ we denote the sign function.  Of course $\ddel_{SI}(\tau,\tau') = - \deld_{SI}(\tau,\tau')$.

\section{Perturbative expansion and heat kernel coefficients}
\label{B}
Here we compute the heat kernel coefficients given in \eqref{hkc} by evaluating the path integral in \eqref{eq:11}.
We present it as a review of worldline methods and to exemplify the equivalence
of the DBC and SI methods for treating the zero mode on the circle \cite{Fliegner:1997rk}.

Let us first rewrite \eqref{eq:11} by factoring out the zero mode integration, and  set up the perturbative expansion 
\be \begin{aligned}
{\rm Tr} \left[ \sigma(x)\, e^{-\beta {\cal R}}\right] & =
\int_{_{PBC}} Dx \, \left(\frac{1}{\beta} \int_0^\beta dt \,  \sigma(x(t))\right)e^{-S[x]} \\
&=\int d^D x_0 \int Dq \, \left(\int_0^1 d\tau \,  \sigma(x_0+q(\tau))\right)e^{-S[x_0+q]} \\
&= \int \frac{d^D x_0}{\left(2\pi\beta\right)^\frac{D}{2}}
\ \Big \la \int_0^1 d\tau\, \sigma(x_0+q(\tau))\, e^{-S_{int}[x_0+q]}\Big\ra
\end{aligned}
\label{pi-sdw}
\ee
where we have rescaled the time in the insertion and action, with the latter taking the form
\be 
S[x_0+q] = S_0[q] + S_{int}[x_0+q] = \frac{1}{\beta}\int_0^1 d\tau\, \frac{1}{2} \dot{q}^2(\tau) 
+ \beta  \int_0^1 d\tau\, V(x_0+q(\tau)) \;.
\ee
Normalized averages with the free path integral are denoted by angle brackets, $\la1\ra=1$, 
and we have extracted  the overall normalization constant 
$\int Dq\, e^{-S_0[q]} = \left(2\pi\beta\right)^{-\frac{D}{2}} $.

The perturbative expansion is implemented by Taylor expanding about $x_0$ the function $\sigma(x)$ 
and the potential  $V(x)$,  and further expanding the exponential of the interaction term. 
Next one computes the correlation functions by Wick contractions. 
Keeping exponentiated the terms that generate disconnected diagrams,
and recalling that the propagators carry a factor of $\beta$,
 we find at  order $\beta^3$
\ba
& 
\Big \la \int_0^1 d\tau\, \sigma(x_0+q) \, e^{-S_{int}[x_0+q]}\Big\ra  
\\
&=
e^{-\beta V(x_0)}  
\biggl[ \sigma(x_0)  
 \biggl ( 1+ \frac{\beta^2}{2}\pa^2 V (x_0)\,  {\bf J}_1 
- \frac{\beta^3}{8}\pa^4 V (x_0)\,  {\bf J}_2 
- \frac{\beta^3}{2}\pa_\mu V  (x_0) \pa^\mu V (x_0) \, {\bf J}_3
\biggr) 
\\
&\qquad  \qquad  \quad  
+ \pa_\mu \sigma (x_0)  \biggl ( \beta^2 \pa^\mu V (x_0)  \, {\bf J}_3
- \frac{\beta^3}{2}\pa^\mu \pa^2V (x_0) \, {\bf J}_4 \biggr) 
\\
&\qquad  \qquad  \quad  
+ \pa^2 \sigma (x_0)  \biggl ( -  \frac{\beta}{2} \,  {\bf J}_1\biggr) 
\biggl ( 1+  \frac{\beta^2}{2} \pa^2 V(x_0) \,  {\bf J}_1  \biggr) 
+ \pa_\mu \pa_\nu \sigma (x_0)  \pa^\mu \pa^\nu V (x_0)  
\biggl ( -  \frac{\beta^3}{2} \,  {\bf J}_5 \biggr) 
\\
&\qquad  \qquad  \quad  
+ \pa_\mu \pa^2 \sigma (x_0) \pa^\mu V (x_0) \biggl ( -  \frac{\beta^3}{2} \,  {\bf J}_4 \biggr) 
\\
&\qquad  \qquad  \quad  
+ \pa^4 \sigma (x_0) \frac{\beta^2}{8}   \, {\bf J}_2 \biggr ] + O(\beta^4)
\label{b3}
 \end{aligned} \ee
where one should expand the exponential in front by keeping only the powers of $\beta$
needed to match the chosen perturbative order. 
The ${\bf J}$'s denote the worldline Feynman diagrams,  
where lines depict propagators and dots denote vertices which include an integration over the time $\tau\in[0,1]$.
They are as follows
\be 
\begin{aligned}
{\bf J}_1 
&=\OneVertexBubble = \int_0 \del_{00} = 
\begin{cases}
-\frac{1}{6} & DBC\\
-\frac{1}{12} & SI
\end{cases}
\\
\\
{\bf J}_2 &=
\OneVertexTwoBubbles = \int_{0} \del^2_{00} = \begin{cases}
\frac{1}{30} & DBC\\
\frac{1}{144} & SI
\end{cases}
\\
\\
{\bf J}_3 &=
\Propagator = \int_{01}  \del_{01} = \begin{cases} 
-\frac{1}{12} & DBC\\
0 & SI
\end{cases}
\\
\\
{\bf J}_4 &=
\PropagatorBubble = \int_{01} \del_{01}\del_{11} = \begin{cases}
\frac{1}{60} & DBC\\
0 & SI
\end{cases}
\\
\\
{\bf J}_5 &=
\TwoVertexOneBubble = \int_{01}\del^2_{01} =  \begin{cases}
\frac{1}{90} & DBC \\
\frac{1}{720} & SI \;.
\end{cases}
\ea
We have given their values both in  DBC and SI. To verify explicitly the equivalence
between DBC and SI, we first plug in 
\eqref{b3} in \eqref{pi-sdw}, perform an integration by parts to free the function $\sigma$ from derivatives, 
and drop total derivative terms. Comparing with \eqref{5}  we recognize the following 
Seeley-DeWitt coefficients
\ba
&a_0(x) = 1 \\
&a_1(x)  = - V \\
&a_2(x) =  \frac12 V^2 + ({\bf J}_1-{\bf J}_3) \pa^2 V 
 \\
&a_3(x) =  -\frac16  V^3 + ({\bf J}_3-{\bf J}_1) V\partial^2 V 
+ \frac12 ({\bf J}_3-{\bf J}_1) \partial_\mu V \partial^\mu V 
- \frac14 ( {\bf J}_2+  {\bf J}_1^2 - 4{\bf J}_4 + 2 {\bf J}_5) \partial^4 V 
\ea
which reproduce those quoted in \eqref{hkc}, independently of the propagator used.
Note that the manifest translational invariance of the SI method allows to get rid of one of the time integrations
in Feynman diagrams. One may use it to fix $\tau=0$ in the insertion, thus relating \eqref{eq:11} to \eqref{8}.

Alternatively, one could compute eq. \eqref{8}  directly with the  DBC method. 
The answer is encoded in the second  line of eq. \eqref{b3} 
(the one proportional to $\sigma(x_0)$), from which one extracts the expected answer.

\section{Evaluation of $b_3$}\label{C}

Here we give additional details on the evaluation of $b_3$.
First we list the diagrams needed for evaluating the leading term discussed in section \ref{five}.
The worldline diagrams have been computed both in DBC and SI, which serves as a check 
on the final result
\ba
{\bf I}_1&= \int_{01}\del^2_{00}\ddel_{01}\deld_{01} = \begin{cases}
-\frac{1}{280},DBC\\
-\frac{1}{1728},SI
\end{cases}\\
{\bf I}_2&= \int_{01}\ddel_{01}\deld_{01}\del_{01}\del_{11} = \begin{cases}
-\frac{1}{420},DBC\\
-\frac{1}{4320},SI
\end{cases}\\
{\bf I}_3&= \int_{01}\ddel_{00}\deld_{01}\del^2_{11} = \begin{cases}
-\frac{1}{1260},DBC\\
0,SI
\end{cases}\\
{\bf I}_4&= \int_{01}\ddel_{00}\del_{01}\ddel_{11}\del_{11} = \begin{cases}
\frac{1}{5040},DBC\\
0,SI
\end{cases}\\ 
{\bf I}_5&= \int_{01}\del_{00}\ddel_{01}\ddel_{11}\del_{11} = \begin{cases}
-\frac{1}{2520},DBC\\
0,SI
\end{cases}\\
{\bf I}_{6}&= \int_{01}\ddel_{00}\deld_{01}\del_{11}\del_{01} = \begin{cases}
-\frac{1}{5040},DBC\\
0,SI
\end{cases}\\ 
{\bf I}_{7}&= \int_{01}\del_{00}\ddel_{01}\deld_{01}\del_{11} = \begin{cases}
-\frac{17}{5040},DBC\\
-\frac{1}{1728},SI
\end{cases}\\
{\bf I}_{8}&= \int_{01}\ddel_{01}\deld_{01}\del^2_{01} = \begin{cases}
-\frac{1}{560},DBC\\
-\frac{11}{60480},SI \;.
\end{cases}
\ea

Let us now consider the other contributions. 
We made extensively use of the color trace and partial integration (both in spacetime and on the worldline), as they
allow to collect identical Wick contractions.
We start by considering the terms arising from the expansion of  two vertices  $S_{int}$ 
coupling $F_{\mu\nu}$ to $V$ 
\begin{align}\label{C1}
C_1 (x_0)&=-\frac{1}{16}\int_{01}\la 
\dot{q}^\mu_0 q^\nu_0 \dot{q}^\alpha_1 q^\beta_1 q^\gamma_1 q^\delta_1
\ra  G_{\nu\mu} (x_0) \nabla_\delta\nabla_\gamma F_{\beta \alpha}(x_0) V (x_0) \\
\label{C2}
C_2(x_0)&=-\frac{1}{8}\int_{012}\la \dot{q}^\mu_0 q^\nu_0 \dot{q}^\alpha_1 q^\beta_1
q^\gamma_2 q^\delta_2 \ra G_{\nu\mu}(x_0)F_{\beta\alpha}(x_0)\nabla_\delta\nabla_\gamma V(x_0)\\
\label{C3}
C_3(x_0)&=-\frac{1}{6}\int_{012}\la \dot{q}^\mu_0 q^\nu_0 \dot{q}^\alpha_1 q^\beta_1
q^\gamma_1 q^\delta_2 \ra G_{\nu\mu}(x_0)\nabla_\gamma F_{\beta\alpha}(x_0)\nabla_\delta V 
(x_0)\\
\label{C4}
C_4(x_0)&=-\frac{1}{9}\int_{01}\la \dot{q}^\mu_0 q^\nu_0 q^\rho_0\dot{q}^\alpha_1  q^\beta_1 q^\gamma_1
\ra \partial_\rho G_{\nu\mu}(x_0) \nabla_\gamma F_{\beta\alpha}(x_0) V(x_0)\\
\label{C5}
C_5(x_0)&=-\frac{1}{16}\int_{01}\la 
\dot{q}^\mu_0 q^\nu_0 q^\rho_0 q^\sigma_0 \dot{q}^\alpha_1 q^\beta_1
\ra \partial_\rho\partial_\sigma G_{\nu\mu}(x_0) F_{\beta\alpha}(x_0) V(x_0)\\
\label{C6}
C_6(x_0)&=-\frac{1}{6}\int_{012}\la \dot{q}^\mu_0 q^\nu_0 q^\rho_0 \dot{q}^\alpha_1 q^\beta_1 q^\gamma_2\ra \partial_\rho G_{\nu\mu}(x_0) F_{\beta\alpha}(x_0)\nabla_\gamma V(x_0)
\end{align}
where, strictly speaking,  the derivatives in the  Taylor expansion of the potential are standard derivative, 
but we have covariantized them anticipating the effect of the time ordering with insertions
of vertices with bare $A_\mu$, as discussed in sec. \ref{four} (see the example in \eqref{comm}).
In addition, we have the terms with three vertices and containing the scalar potential 
\begin{align}
C_7(x_0)&= 
\frac{1}{8}\int_{0123}\la \dot{q}^\alpha_0 q^\beta_0 \dot{q}^\gamma_1 q^\delta_1 \dot{q}^\mu_3 q^\nu_3\ra g_{012}
\, G_{\nu\mu}(x_0)F_{\beta\alpha}(x_0)F_{\delta\gamma}(x_0)V(x_0) \label{C7}\\
C_8(x_0)&=\frac{\beta}{8} \int_{01}\la \dot{q}^\mu_0 q^\nu_0 \dot{q}^\alpha_1 q^\beta_1\ra 
\,G_{\nu\mu}(x_0)F_{\beta\alpha}(x_0)V^2(x_0)
\label{C8}
\end{align}
where the function $g_{012}$ contains step functions that take care of the time ordering
\ba
g_{012}&=\theta_{01}\theta_{12}+\theta_{12}\theta_{20}+\theta_{20}\theta_{01}\;.
\ea
In $C_8$ the time ordering is not necessary, thanks to the color trace. It is then recognized as
a disconnected diagram that is calculated straightforwardly from previous results. 
Moving on to the term with three non-abelian field strengths we find
\be \label{D}
C_9(x_0) = \frac{1}{16\beta}\int_{0123}\la\dot{q}^\mu_0 q^\nu_0 \dot{q}^\alpha_1 q^\beta_1 \dot{q}^\gamma_2 q^\delta_2 \dot{q}^\epsilon_3 q^\eta_3 \ra \theta_{12}\theta_{23}G_{\nu\mu}(x_0)F_{\beta\alpha}(x_0)F_{\delta\gamma}(x_0)F_{\eta\epsilon}(x_0) \;.
\ee 
 
Let us now evaluate these terms. For $C_1$ we get
\ba
C_1 
=-&\frac{\beta^3}{4}G^{\mu\nu}\nabla^2 F_{\mu\nu}V 
\,{\bf H}_{1}+\frac{\beta^3}{4}G^{\mu\nu}F_{\nu\alpha}F^\alpha{}_\mu V\,{\bf H}_{1} 
\ea
For $C_2$ we get
\ba
C_2
&=\frac{\beta^3}{2}\left(-\partial^2 G^{\mu\nu}F_{\mu\nu}-G^{\mu\nu}\nabla^2 F_{\mu\nu} -\partial^\alpha G^{\mu\nu}\nabla_\alpha F_{\mu\nu} -2\partial_\alpha G^{\alpha\nu}\nabla^\mu F_{\mu\nu}	+2G^{\alpha\nu}F_{\nu\mu}F^\mu{}_\alpha 	\right)V\, {\bf K}_{1}\\
&-\frac{\beta^3}{4}\left(2\partial^\alpha G^{\mu\nu}\nabla_\alpha F_{\mu\nu} +\partial^2G^{\mu\nu}F_{\mu\nu}+G^{\mu\nu}\nabla^2 F_{\mu\nu}		\right)V\, {\bf F}_3\, {\bf J}_1
\ea
with the new diagram
\be
{\bf K}_{1}=\int_{012}\ddel_{01}\ddel_{12}\del_{02}=\begin{cases}
\frac{1}{360}, DBC\\
\frac{1}{720}, SI \;.
\end{cases}
\ee
Next, $C_3$ evaluates to
\ba
C_3 &= \frac{\beta^3}{2}
\left(G^{\mu\nu}\nabla^2 F_{\mu\nu}  
+ \partial^\sigma G^{\mu\nu}\nabla_\sigma F_{\mu\nu}
 \right)V
\, {\bf K}_{2}\\
&+\frac{\beta^3}{2}
\left( G^{\mu\nu}\nabla^2 F_{\mu\nu} +2
\partial_\alpha G^{\alpha\mu}\nabla^\sigma F_{\sigma\mu}
-4 G^{\alpha\mu}F_{\mu\sigma}F^\sigma{}_\alpha		\right)V \, {\bf K}_{3}
\ea
with 
\ba
{\bf K}_{2}&=\int_{012}\ddel_{01}\deld_{01}\del_{12}=\begin{cases}
\frac{1}{120}, DBC\\
0, SI
\end{cases}\\
{\bf K}_{3}&=\int_{012}\ddel_{01}\ddel_{11}\del_{02}=\begin{cases}
\frac{1}{720}, DBC\\
0, SI
\end{cases}
\ea
while  \eqref{C4} and \eqref{C5} produce
\ba
C_4
&= -\frac{\beta^3}{2}\partial^\nu G_{\nu\mu}\nabla_\alpha F^{\alpha\mu}V{\bf H}_4  - \frac{\beta^3}{2}\partial^\lambda G^{\mu\nu}\nabla_\lambda F_{\mu\nu}V {\bf H}_2 \\
C_5 &= -\frac{\beta^3}{4}\partial^2 G^{\mu\nu}F_{\mu\nu}V \,{\bf H}_1\;.
\ea
For $C_6$ we have 
\ba 
C_6(x_0) 
&= \frac{\beta^3}{2}
\left(\partial^2 G^{\mu\nu}F_{\mu\nu} +\partial^\sigma G^{\mu\nu}\nabla_\sigma F_{\mu\nu}		\right)V \, {\bf K}_{2}
+\frac{\beta^3}{2}
\left(\partial^2 G^{\mu\nu}F_{\mu\nu}	+
2\partial_\lambda G^{\lambda\mu}	\nabla^\alpha F_{\alpha\mu} 
\right) V\,
{\bf K}_{3} \;.
\ea
Finally, we consider $C_7$ that mixes with the above terms   
\ba 
C_7 &=\frac{\beta^3}{4}G^{\rho\nu}F_{\nu\mu}F^\mu{}_\rho V \left({\bf K}_{4} -{\bf K}_{5} +{\bf K}_{6} - {\bf K}_{7}\right)
\ea
and contains the following integrals 
\ba
{\bf K}_{4}&=\int_{0123}\ddel_{03}\deld_{01}\deld_{13}\,g_{012}=\begin{cases}
\frac{1}{240}, DBC\\
\frac{1}{360}, SI
\end{cases}\\
{\bf K}_{5}&=\int_{0123}\ddeld_{13}\ddel_{03}\del_{01}\,g_{012}=\begin{cases}
\frac{7}{720}, DBC\\
\frac{1}{360}, SI
\end{cases}\\
{\bf K}_{6}&=\int_{0123}\ddeld_{01}\deld_{03}\del_{13}\,g_{012}=\begin{cases}
\frac{1}{720}, DBC\\
\frac{1}{720}, SI
\end{cases}\\
{\bf K}_{7}&=\int_{0123}\ddel_{01}\deld_{03}\ddel_{13}\,g_{012}=\begin{cases}
-\frac{1}{240}, DBC\\
-\frac{1}{360}, SI \;.
\end{cases}\\
\ea
Collecting all the terms from  $C_1$ to $C_7$
we find that they sum up to the entire second line of $b_3$ in \eqref{ghkc}, independently of the propagator used.
We have performed integrations by parts to reduce to a set of independent terms
(in particular, we have left $V$ free from derivatives).

As for $C_8$, since the time ordering can be neglected, we find that
it  is given by a disconnected  correlation
functions that embeds the first piece of $b_2$ and produces
\be
C_8 = -\frac{\beta^3}{4}G^{\mu\nu}F_{\mu\nu}V^2\, {\bf F}_{3} = 
\frac{\beta^3}{48}G^{\mu\nu}F_{\mu\nu}V^2 
\ee
which sits inside \eqref{ghkc} as the last term of the first row of  $b_3$.

Finally, we consider \eqref{D}, that contains three non-abelian field strengths $F_{\mu\nu}$.
In  \eqref{D} we kept the time ordering encoded in the step functions.
However, one may note that the Wick contractions produce terms that have no ordering ambiguities under the trace,
implying that  the abelian limit contains precisely the same information. 
Thus, we are allowed to drop the step functions and consider the equivalent (under the color trace) form
\be 
\tilde{C_9}(x_0) =\frac{1}{96\beta}\int_{0123}\la \dot{q}^\mu_0 q^\nu_0 \dot{q}^\alpha_1 q^\beta_1 \dot{q}^\gamma_2 q^\delta_2 \dot{q}^\epsilon_3 q^\eta_3		\ra G_{\nu\mu}(x_0)F_{\beta\alpha}(x_0)F_{\delta\gamma}(x_0)F_{\eta\epsilon}(x_0)\;.
\ee
We compute it as
\be 
\tilde{C_9} = \frac{\beta^3}{8} G^{\mu\nu}F_{\mu\nu}F^2\,  ({\bf F}_3)^2 +\frac{\beta^3}{2}
G^{\mu\nu}F_{\nu\rho}F^{\rho\lambda}F_{\lambda\mu}\,
{\bf K}_{8}	
\ee
where the first term arises from disconnected diagrams with 
${\bf F}_3=-\frac{1}{12}$ already  given in \eqref{f3}, 
and 
\be
{\bf K}_{8} = \int_{0123}\ddel_{01}\ddel_{12}\ddel_{23}\ddel_{30} = 
\frac{1}{720}   
\ee 
valid in DBC and SI, 
thus producing the second and third term of $b_3$ in \eqref{ghkc}.

\end{document}